\documentclass[11pt,a4paper]{article}
\usepackage{jheppub}

\usepackage{epsfig,multicol}
\usepackage{amsmath,epsfig}
\usepackage{amsfonts}
\usepackage{multirow}
\usepackage{mathrsfs}
\usepackage{graphicx}
\usepackage{amsmath}
\usepackage{amssymb}
\usepackage{bm}
\usepackage{bbm}
\usepackage{diagbox}
\usepackage{ulem}
\usepackage{slashbox}

\def\pslash{p\!\!\!\slash}
\def\pbarslash{\bar{p}\!\!\!\slash}

\def\Pslash{P\!\!\!\slash}

\def\OMIT#1{}

\newcommand{\nn}{\nonumber}

\newcommand{\beq}{\begin{equation}}
\newcommand{\eeq}{\end{equation}}
\newcommand{\bqa}{\begin{eqnarray}}
\newcommand{\eqa}{\end{eqnarray}}

\newcommand\fverb{\setbox\fverbbox=\hbox\bgroup\verb}
\newcommand\fverbdo{\egroup\medskip\noindent%
			\fbox{\unhbox\fverbbox}\ }
\newcommand\fverbit{\egroup\item[\fbox{\unhbox\fverbbox}]}
\newbox\fverbbox

\title{\mbox{}\\[10pt]
Next-to-Next-to-Leading-Order Radiative Corrections to $e^+e^-\to\chi_{cJ}+\gamma$ at $B$ factory}
\author[a]{Wen-Long Sang,}
\author[b,c]{Feng Feng}
\author[c,d]{Yu Jia}

\affiliation[a]{School of Physical Science and Technology, Southwest University, Chongqing 400700, China}
\affiliation[b]{China University of Mining and Technology, Beijing 100083, China}
\affiliation[c]{Institute of High Energy Physics and Theoretical Physics Center for Science Facilities, Chinese Academy of Sciences, Beijing 100049, China}
\affiliation[d]{School of Physics, University of Chinese Academy of Sciences, Beijing 100049, China}

\emailAdd{wlsang@swu.edu.cn}
\emailAdd{F.Feng@outlook.com}
\emailAdd{jiay@ihep.ac.cn}

\abstract{
Within the nonrelativistic QCD (NRQCD) factorization framework, we have computed the ${\mathcal O}(\alpha_s^2)$ corrections to
the exclusive production of $P$-wave spin-triplet charmonia $\chi_{cJ}(J=0,1,2)$ accompanied with a hard photon at $B$
factory. For the first time, we have explicitly verified the validity of NRQCD factorization
for exclusive $P$-wave quarkonium production to two-loop order.
Unlike the $\chi_{cJ}$ electromagnetic decays, the $\mathcal{O}(\alpha_s^2)$ corrections
are found to be smaller than the $\mathcal{O}(\alpha_s)$ corrections in all three channels  $e^+e^-\to \chi_{c0,1,2}+\gamma$.
In particular, the ${\mathcal O}(\alpha_s^2)$ corrections appear moderate for $\chi_{c1}+\gamma$ case,
and marginal for $\chi_{c0}+\gamma$. Moreover, the predictions in next-to-next-to-leading order (NNLO) accuracy
for the production rates of $\chi_{c0,1}+\gamma$ are insensitive to the renormalization and factorization scales.
All these features may indicate that perturbative expansion in these two channels exhibits a decent convergence behavior.
By contrast, both the ${\mathcal O}(\alpha_s)$ and ${\mathcal O}(\alpha_s^2)$ corrections to the
$\chi_{c2}+\gamma$ production rate are sizable, which reduce the Born order cross section
by one order of magnitude after including the NNLO perturbative corrections.
Taking the values of the long-distance NRQCD matrix elements from nonrelativistic potential model, our
prediction to $\chi_{c1}+\gamma$ production rate is consistent with the recent {\tt Belle} measurement.
The NNLO predictions to the $\chi_{c0,2}+\gamma$ production rates are much smaller than that for $\chi_{c1}+\gamma$,
which seems to naturally explain why the $e^+e^-\to \chi_{c0,2}+\gamma$ channels have escaped experimental detection to date.
}
\begin{document}
\maketitle
\section{Introduction\label{sec:intro}}

Heavy quarkonia, the tightly-bound systems composed of a heavy quark and a heavy antiquark,
are generally viewed as the simplest hadrons in Quantum Chromodynamics (QCD).
A peculiar trait of quarkonia is the coexistence of several distinct mass scales,
which makes it an interesting and unique laboratory to sharpen our understanding about the
interplay between perturbative and nonperturbative aspects of QCD.
Since the heavy (anti)quark inside a quarkonium is essentially nonrelativistic,
the mainstream theoretical method is firmly based on the modern effective field theory (EFT) doctrine,
the so-called Nonrelativistic QCD (NRQCD) factorization approach~\cite{Bodwin:1994jh}.
This approach allows to systematically disentangle the short-distance and long-distance effects,
formalized by a double expansion in QCD strong coupling $\alpha_s$ and heavy quark velocity $v$.
In the past two decades, NRQCD factorization has been widely employed to tackle innumerable
quarkonium production and decay processes.

Thanks to its enormous luminosity and simplicity of the initial state,
$B$ factories have acted as an fertile and clean playground to investigate charmonium production.
For instance, in the past two decades, there have emerged a handful of experimental measurements about exclusive charmonium production processes~\cite{Abe:2002rb,Aubert:2005tj,Aubert:2001pd,Abe:2001za,Pakhlov:2009nj}, together with intensive theoretical investigations using NRQCD approach~\cite{Braaten:2002fi,Liu:2002wq,Hagiwara:2003cw,Ma:2008gq,Gong:2009kp}
(for a comprehensive list of references, we refer the interested readers to the recent review article~\cite{Brambilla:2014jmp}).

Among various charmonium production processes, the exclusive production of positive-$C$-parity charmonium associated with a hard photon,
{\it e.g.}, $e^+e^-\to \eta_c(\chi_{cJ})+\gamma$ ($J=0,1,2$), is of special interest.
Due to their simplicity, these processes can be regarded as the golden channels to test our understanding of charmonium production mechanism
and the utility of NRQCD approach. The leading-order (LO) cross section for $e^+e^-\to \eta_c(\chi_{c0,1,2})+\gamma$ at $B$ factory
was predicted in \cite{Chung:2008km}. The next-to-leading-order (NLO) perturbative corrections were subsequently computed
in \cite{Sang:2009jc,Li:2009ki}, where the ${\cal O}(\alpha_s)$ corrections in the $\chi_{c2}+\gamma$ channel turn out to be sizable,
even exceeding  $-60\%$~\footnote{When $\sqrt{s}\gg m_c$, the collinear logarithm $\ln s/m_c^2$ in NRQCD short-distance coefficients
can get large, which may potentially ruin the convergence of fixed-order perturbative expansion.
For $e^+e^-\to \eta_c+\gamma$, there have been attempts to resum the leading logarithms~\cite{Jia:2008ep} and next-to-leading logarithms~\cite{Chung:2019ota} to all orders in $\alpha_s$.}.
On the other hand, since charm quark is not decently heavy, one may expect that relativistic corrections might also have
important impact. The leading relativistic corrections to $e^+e^-\to \eta_c+\gamma$ were first considered in \cite{Sang:2009jc}.
The relativistic corrections to $e^+e^-\to \chi_{c0,1,2}+\gamma$ were first explored in \cite{Xu:2014zra}, yet missing the contribution
due to the NRQCD operators that explicitly contains the chromoelectric field.
Very recently, the complete $\mathcal{O}(v^2)$ corrections to these $P$-wave charmonium exclusive production processes have been given in \cite{Brambilla:2017kgw}.
Unfortunately, the values of various $\mathcal{O}(v^2)$ NRQCD long-distance matrix elements (LDMEs) are poorly constrained,
therefore it is difficult to present accurate predictions for the $\chi_{cJ}+\gamma$ production rates.

Leaving relativistic corrections aside, one has witnessed remarkable progress in deducing the higher-order perturbative
corrections for various quarkonium decay and production processes. More than two decades ago, the next-to-next-to-leading order (NNLO)
perturbative corrections for the simplest quarkonium electromagnetic decay, $\Upsilon(J/\psi)\to e^+e^-$,
were analytically determined~\cite{Czarnecki:1997vz,Beneke:1997jm}. It is worth noting that, the ${\cal O}(\alpha_s^3)$ corrections for these channels have also been available recently~\cite{Marquard:2014pea,Beneke:2014qea}.
In recent years, with the advance of numerical and analytic multi-loop technology, a number of NNLO perturbative corrections to
more difficult quarkonium decay processes have been accomplished, {\it e.g.},
$\eta_{b,c}\to\gamma\gamma$~\cite{Czarnecki:2001zc,Feng:2015uha}, $B_c\to \ell \nu$~\cite{Onishchenko:2003ui,Chen:2015csa},
$\chi_{c0,2}\to\gamma\gamma$~\cite{Sang:2015uxg}, and
$\eta_{c,b}\to {\rm light\;hadrons}$~\cite{Feng:2017hlu}.
For most of the aforementioned processes, the NNLO radiative corrections turn out to be sizable and
significantly modify the lower-order NRQCD predictions, especially for charmonia.

The NNLO perturbative corrections to the simplest channel of exclusive  quarkonium production,
the $\gamma\gamma^*\to \eta_{c,b}$ transition form factor, have also been reported recently~\cite{Feng:2015uha,Wang:2018lry}.
Very recently, the NNLO radiative corrections to the famous double-charmonium production process at $B$ factory,
$e^+e^-\to J/\psi+\eta_c$, have also been inferred~\cite{Feng:2019zmt}.
In this case, the ${\cal O}(\alpha_s^2)$ corrections are observed to have moderate effect.
It is encouraging that the state-of-the-art NRQCD prediction is consistent with the
{\tt BaBAR} measurement~\cite{Aubert:2005tj}.
Moreover, the NNLO radiative corrections to $e^+e^-\to \eta_c+\gamma$ has also recently been computed analytically at lowest order in $v$~\cite{Chen:2017pyi}, again with moderate impact~\footnote{Note that the NRQCD short-distance coefficient has a rather
lengthy expressions in term of Goncharov polylogarithms, the integrals over polylogarithms and complete elliptic integrals.}.
Very recently, the NNLO corrections to this process were also reinvestigated with the renormalization scales chosen by the principle of
maximum conformality~\cite{Yu:2020tri}.

In this work, we proceed to further evaluate the NNLO perturbative corrections for $e^+e^-\to \chi_{cJ}+\gamma$ at $B$ factory.
This work constitutes the first NNLO perturbative correction calculation for the $P$-wave quarkonium production process.
It is of theoretical curiosity to examine the validity of NRQCD factorization framework for this case.
We also wish to examine the convergence of perturbative expansion in this channel, as well as
confront our predictions with the latest measurement on $e^+e^-\to \chi_{c1}+\gamma$ by {\tt Belle} Collaboration~\cite{Jia:2018xsy}.

The remainder of this paper is structured as follows.
In section~\ref{sec-theory}, after outlining the NRQCD factorization formula for $e^+e^-\to \chi_{cJ}+\gamma$,
we describe the theoretical strategy to deduce the NRQCD SDCs associated with the $J=0,1,2$ channels.
In section~\ref{sec-calculation}, we first briefly describe some technicalities encountered in two-loop calculation,
then present the numerical results for various NRQCD short-distance coefficients.
Section~\ref{sec-phen} is devoted to the phenomenological analysis and discussion.
We summarize in section~\ref{summary}.

\section{Theoretical background for $P$-wave onium exclusive production   \label{sec-theory}}

In accordance with the NRQCD factorization ansatz,
the production rates of $e^+e^-\to \chi_{cJ}+\gamma$ can be expressed in the following
factorized form:
\begin{eqnarray}\label{eq-NRQCD-formula}
\sigma(\chi_{cJ}+\gamma)
 &=& F_1(^3P_J)
\langle \mathcal{O}(^3P_J)\rangle+{\mathcal O}(\sigma v^2),
\end{eqnarray}
where $F_1(^3P_J)$ ($J=0,1,2$) represent the corresponding NRQCD short-distance coefficients (SDCs).
Owing to asymptotic freedom, these coefficients can be computed in perturbation theory, order by order in powers of
the strong coupling constant $\alpha_s$.
$\langle\mathcal{O}(^3P_J)\rangle$ represent the process-independent NRQCD long-distance matrix elements (LDMEs),
which bear a genuinely nonperturbative origin, and are defined as
\begin{eqnarray}\label{eq-NRQCD-operators}
\langle \mathcal{O}(^3P_J) \rangle \equiv \left|\langle \chi_{cJ}|
\psi^{\dagger}{\cal K}_{^3P_J} \chi |0\rangle\right|^2,
\end{eqnarray}
where $\psi$ and $\chi^\dagger$ denote the Pauli spinor fields annihilating a heavy quark and antiquark, respectively, and
\begin{subequations}
\begin{eqnarray}
{\cal K}_{^3P_{0}}&=&\frac{1}{\sqrt{3}}\left(-\frac{i}{2}\overleftrightarrow{{\bf
D}}\cdot {\bm\sigma}\right),\\
{\cal K}_{^3P_1}&=&\frac{1}{\sqrt{2}}\left(-\frac{i}{2}\overleftrightarrow{{\bf
D}}\times\bm{\sigma}\right),\\
{\cal K}_{^3P_2}&=&-{i\over 2}\overleftrightarrow{D}^{(i}\sigma^{j)}.
\end{eqnarray}
\end{subequations}
Note that the polarization indices affiliated with $\chi_{c1,2}$ have not been summed in \eqref{eq-NRQCD-operators}.

Invoking the approximate heavy quark spin symmetry, we have the following simplifying relations:
\begin{eqnarray}\label{NRQCD:heavy:quark:spin:sym}
\langle\mathcal{O}(^3P_0)\rangle\approx
\langle\mathcal{O}(^3P_1)\rangle\approx \langle\mathcal{O}(^3P_2)\rangle.
\end{eqnarray}

Since NRQCD SDCs are insensitive to the nonperturbative hadronization effects, they can be deduced with the aid of the standard perturbative matching technique.
That is, by replacing the physical $\chi_{cJ}$ meson with a fictitious onium composed of free $c\bar{c}$ pair, carrying the quantum number ${}^3P_J$,
we compute both sides of \eqref{eq-NRQCD-formula} order by order in $\alpha_s$.
After this replacement, \eqref{eq-NRQCD-formula} becomes
\begin{eqnarray}\label{eq-NRQCD-formula-pert}
\sigma(c\bar{c}(^3P_J)+\gamma)
 =F_1(^3P_J) \langle\mathcal{O}(^3P_J)\rangle_{c\bar{c}(^3P_J)},
\end{eqnarray}
where the subscript $c\bar{c}(^3P_J)$ in the NRQCD LDMEs indicates that the hadronic states $\chi_{cJ}$ have been replaced by a $c\bar{c}(^3P_J)$ pair,
which can be accessed in perturbation theory. After computing both sides of \eqref{eq-NRQCD-formula-pert} in perturbative QCD and NRQCD,
we are able to solve for the desired NRQCD SDCs order by order in $\alpha_s$. It is worth emphasizing that, for a quarkonium hard
exclusive reaction like in our case,
the factorization \eqref{eq-NRQCD-formula} actually also holds at the amplitude level.

To facilitate the perturbative calculation, we assign the momenta of the $c$ and $\bar{c}$ quarks to be
 \begin{subequations}\label{kinematics-momenta}
 \bqa
p&=&\frac{P}{2}+q, \\
\bar{p}&=&\frac{P}{2}-q,
\eqa
\end{subequations}
where $P$ and $q$ denote the total momentum of the $c\bar{c}$ pair and the relative momentum, respectively.
The on-shell condition $p^2=\bar{p}^2=m^2$ (with $m$ signifying the charm quark mass) enforces that
\bqa
\label{kinematics-on-shell}
&&P\cdot q=0,\qquad P^2=4E^2,
\eqa
with $E=\sqrt{m^2-q^2}\ge m$. Since we are only concerned with the SDCs at the lowest order in $v$,
it is legitimate to approximate the square of the invariance mass of the $c\bar{c}$ pair by $4m^2$.

It is convenient to employ the covariant spin-projector to enforce the $c\bar{c}$ pair in
the spin-triplet state.
The relativistically normalized color-singlet/spin-triplet projector reads~\cite{Bodwin:2013zu}:
\bqa
\label{spin-projector}
\Pi_1^\mu=\frac{-1}{8\sqrt{2}m^2}(\pbarslash-m)\gamma^\mu(\Pslash+2m)(\pslash+m)\otimes {{\bf 1}_c \over \sqrt{N_c}}.
\eqa

The $c\bar{c}({}^3P_J)$ amplitude can be projected out by differentiating
the colour-singlet/spin-triplet quark amplitude ${\cal A}$ with respect to the relative momentum $q$,
followed by setting $q$ to zero:
 \bqa
\label{p-wave-projector}
\mathcal{A}^{(J)}=\epsilon^{(J)}_{\mu\nu}\frac{d}{dq_\nu}{\rm Tr}[\Pi_1^\mu \mathcal{A}]\Big |_{q=0},
\eqa
with $\epsilon^{(J)}_{\mu\nu}$ denoting the polarization vectors affiliated with
$J=0,1,2$.

In order to obtain the unpolarized cross section, we also need sum over all possible
polarizations for each $J$. It is convenient to employ the polarization sum
identities given in \cite{Petrelli:1997ge}:
\begin{subequations}
\label{polarization-sum}
 \bqa
\epsilon^{(0)}_{\mu\nu}\epsilon^{(0)*}_{\alpha\beta}&=&\frac{1}{d-1}\Pi_{\mu\nu}\Pi_{\alpha\beta},\\
\epsilon^{(1)}_{\mu\nu}\epsilon^{(1)*}_{\alpha\beta}&=&\frac{1}{2}\left(\Pi_{\mu\alpha}\Pi_{\nu\beta}-\Pi_{\mu\beta}\Pi_{\nu\alpha}\right),\\
\epsilon^{(2)}_{\mu\nu}\epsilon^{(2)*}_{\alpha\beta}&=&\frac{1}{2}\left(\Pi_{\mu\alpha}\Pi_{\nu\beta}+\Pi_{\mu\beta}\Pi_{\nu\alpha}\right)
-\frac{1}{d-1}\Pi_{\mu\nu}\Pi_{\alpha\beta},
\eqa
\end{subequations}
where $d=4-2\epsilon$ signifies the space-time dimensions, and the polarization tensor $\Pi_{\mu\nu}(P)$
is defined through
 \bqa
\Pi_{\mu\nu}=-g_{\mu\nu}+\frac{P_\mu P_\nu}{4m^2}.
\eqa

Now we have collected all the necessary ingredients to evaluate the quark-level cross sections $\sigma(c\bar{c}(^3P_J)+\gamma)$ in perturbative QCD.
In the meanwhile, the perturbative NRQCD matrix elements $ \langle \mathcal{O}(^3P_J)\rangle_{c\bar{c}(^3P_J)}$ can also be carried out.
It is then straightforward to ascertain the SDCs following the matching procedure.

\section{The cross sections through ${\mathcal O}(\alpha_s^2)$
\label{sec-calculation}}

In this section, we first describe the computational techniques utilized to determine the various two-loop SDCs $F_1(^3P_J)$,
then present their numerical expressions.

\begin{figure}[htbp]
 	\centering
 \includegraphics[width=1.\textwidth]{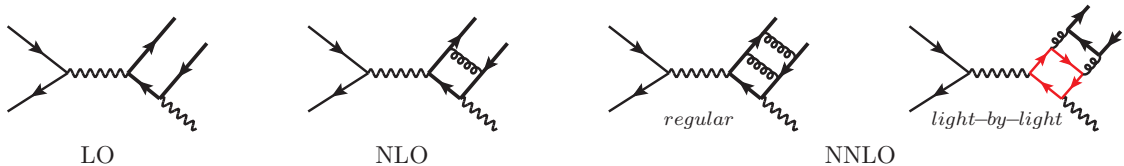}
 \caption{Some representative Feynman diagrams for $e^+e^-\to c\bar{c}({}^3P_J)+\gamma$,
 which are drawn by \texttt{JaxoDraw}~\cite{Binosi:2008ig}
 \label{fig-feynman-diagram}}
 \end{figure}

We use {\tt FeynArts}~\cite{Hahn:2000kx} to generate the Feynman diagrams for $e^+e^-\to c\bar{c}+\gamma$ and corresponding Feynman amplitude through two-loop order in $\alpha_s$.
Some typical Feynman graphs are displayed in Fig.~\ref{fig-feynman-diagram}.
Employing the color-singlet/spin-triplet projector (\ref{spin-projector}), following the recipe as specified in (\ref{p-wave-projector})
to single out the $P$-wave component of the amplitude, we are able to project out the intended $e^+e^-\to c\bar{c}(^3P_J)+\gamma$ amplitude, order by order in $\alpha_s$.
We then employ the packages {\tt FeynCalc}~\cite{Mertig:1990an,Shtabovenko:2016sxi} and {\tt FormLink}~\cite{Feng:2012tk,Kuipers:2012rf} to carry out the polarization sum according to
(\ref{polarization-sum}).

The leading-order (LO) NRQCD SDCs have long been known~\cite{Sang:2009jc}:
\begin{subequations}\label{sdcs-0}
\begin{eqnarray}
{F}_1^{(0)}(^3P_0) &=&\frac{32e_c^4\alpha^3\pi^2(1-3r)^2}{9m^3s^2(1-r)},\\
{F}_1^{(0)}(^3P_1)&=&\frac{64e_c^4\alpha^3\pi^2(1+r)}{3m^3s^2(1-r)},\\
{F}_1^{(0)}(^3P_2)&=&\frac{64e_c^4\alpha^3\pi^2(1+3r+6r^2)}{9m^3s^2(1-r)},
\end{eqnarray}
\end{subequations}
where $\alpha$ represents the electromagnetic coupling constant, $e_c=2/3$ signifies the charge of charm quark,
$s$ corresponds to the squared center-of-mass energy. The dimensionless ratio $r\equiv 4m^2/s$ is constrained to be less than 1.

Once beyond the LO, we adopt the standard shortcut to directly extract the short-distance coefficient, {\it i.e.}, to take the derivative of
the amplitude with respect to $q$ prior to conducting the loop integral, which amounts to directly extracting the contribution from the hard region
in the context of strategy of region~\cite{Beneke:1997zp}.
Throughout the work we employ the dimensional regularization to regularize the UV and IR divergences.
For the loop integrals, we utilize the packages {\tt Apart}~\cite{Feng:2012iq} and {\tt FIRE}~\cite{Smirnov:2014hma} to conduct partial fraction and the
corresponding integration-by-part (IBP) reduction. We end up with 6 one-loop master integrals (MIs) and 174 two loop MIs,
most of which are complex-valued integrals.
For the one-loop MIs, one can readily work out the analytical expression for all the MIs.
We have confirmed the analytic expression of the ${\cal O}(\alpha_s)$ corrections to the cross section,
first accomplished in \cite{Sang:2009jc}.

It becomes much more challenging to deduce the analytical expressions for all the encountered two-loop MIs.
In this work, we are content with high-precision numerical results~\footnote{In principle, one may make use of those two-loop MIs
encountered in a similar process $e^+e^-\to \eta_c+\gamma$~\cite{Chen:2017pyi}.
Nevertheless, as far as phenomenological analysis is concerned, we feel that it suffices to utilize the entirely numerical recipe to tackle these MIs.}.
We use the modified {\tt FIESTA}~\cite{Smirnov:2013eza} to perform sector decomposition (SD) for the two-loop MIs.
For the real-valued MIs, we directly use {\tt CubPack}~\cite{CubPack} to carry out the numerical integration.
In contrast to the application of SD to the Euclidean region, the singularities encountered in the physical region lie
inside, rather than sit on, the integration boundary, which render the integrals hard to be numerically evaluated.
The difficulty can be overcome to a certain extent by deforming the integration contour via the following variable
transformation prior to decomposing the sectors~\cite{Borowka:2012qfa}:
\begin{eqnarray}
\label{contour-deformation}
z_k=x_k-i\lambda_k x_k (1-x_k)\frac{\partial F}{\partial x_k},
\end{eqnarray}
where $F$ denotes the $F$-term in the $\alpha$ parametrization, $\lambda_k$ is some positive number. Actually, the integration efficiency may vary drastically with ${\lambda_k}$. In our calculation, we first choose a set of ${\lambda_k}$ and utilize {\tt CubPack}
to conduct the first-round rough numerical integration.
For those integrals with large estimated errors, we adjust the values of ${\lambda_k}$ and perform the integration with a fixed number of sample points. The operation will be repeated until we find a optimized values of ${\lambda_k}$, which render the integration error endurable. With the new determined ${\lambda_k}$, the integration will be performed once again with the aid of a parallelized integrator {\tt HCubature}~\cite{HCubature} to reach the desired precision.
To further improve efficiency, we interchange the order of the operations for contour deformation, SD and series expansion
in {\tt FIESTA}, namely we delay the transformation (\ref{contour-deformation}) until the end.
For more technical details, we refer the interested readers to Ref.~\cite{Feng:2019zmt}.

To eliminate UV divergences, we carry out the renormalization procedure by implementing the $\mathcal{O}(\alpha_s^2)$ expressions of the on-shell renormalization constants $Z_2$ and $Z_m$
from \cite{Broadhurst:1991fy,Melnikov:2000zc,Baernreuther:2013caa}. The strong coupling constant $\alpha_s$ is renormalized under $\overline{\rm MS}$ scheme. Nevertheless, the renormalized NNLO squared amplitudes are found to
still contain an uncancelled single IR pole.
This symptom is a common feature specific to NRQCD factorization, which have been encountered many times in NNLO perturbative calculations involving
quarkonium. This IR pole can be factored into the NRQCD LDME, so that the NRQCD SDCs become IR finite.
As a consequence, both of the LDMEs and the corresponding two-loop SDCs develop a $\log \mu_\Lambda$
dependence ($\mu_\Lambda$ refers to NRQCD factorization scale), nevertheless their product must be independent of $\mu_\Lambda$.
In fact, from the coefficient of the single IR pole, one can read off the anomalous dimensions associated with the
NRQCD bilinear currents carrying the quantum numbers $^3P_J$ in \eqref{eq-NRQCD-operators}:
\begin{subequations}
\label{anomalous-dimension}
\bqa
\gamma_{\chi_{c0}}&=&-\pi^2\bigg(\frac{C_AC_F}{6}+\frac{2C_F^2}{3}\bigg),\\
\gamma_{\chi_{c1}}&=&-\pi^2\bigg(\frac{C_AC_F}{6}+\frac{5C_F^2}{12}\bigg),\\
\gamma_{\chi_{c2}}&=&-\pi^2\bigg(\frac{C_AC_F}{6}+\frac{13C_F^2}{60}\bigg).
\eqa
\end{subequations}
Reassuringly, these values exactly agree with those predicted from the renormalization group
analysis in velocity NRQCD~\cite{Hoang:2006ty}.

As mentioned before, we will be content with only providing the numerical expressions for various NRQCD SDCs.
We then substitute these results into \eqref{eq-NRQCD-formula} to predict the unpolarized production rates for
$e^+e^-\to \chi_{cJ}+\gamma$, to the prescribed order in $\alpha_s$.
For numerical calculation, we take the $B$ factory center-of-mass energy to be $\sqrt{s}=10.58$ GeV.
We choose two typical values for charm mass, $m = 1.4$ GeV and $1.68$ GeV, which correspond to the one-loop and two-loop charm quark pole mass
The production rates for $e^+e^-\to \chi_{cJ}+\gamma$  through NNLO accuracy are then predicted to be
\begin{subequations}\label{cs-total-mc-1.4}
\begin{eqnarray}
\sigma(\chi_{c0})&=&\sigma^{(0)}(\chi_{c0})\bigg\{1+\frac{\alpha_s}{\pi}(1.9332)+\frac{\alpha_s^2}{\pi^2}\bigg[\frac{1}{4}\beta_0\ln\frac{\mu_R^2}{4m^2}(1.9332)
+2\gamma_{\chi_{c0}}\ln\frac{\mu_\Lambda}{m}\nn\\
&&+\bigg(0.867143(3){ n_H}-1.6338020(7){ n_L}+5.17(4){\rm lbl}-9.020(3)\bigg)
\bigg]\bigg\}, \\
\sigma(\chi_{c1})&=&\sigma^{(0)}(\chi_{c1})\bigg\{1+\frac{\alpha_s}{\pi}(-3.1597)+\frac{\alpha_s^2}{\pi^2}\bigg[\frac{1}{4}\beta_0\ln\frac{\mu_R^2}{4m^2}(-3.1597)
+2\gamma_{\chi_{c1}}\ln\frac{\mu_\Lambda}{m}\nn\\
&&+\bigg(0.037950(1){ n_H}-0.5954237(4){ n_L}-4.191(3){\rm lbl}-17.337(2)\bigg)
\bigg]\bigg\},\\
\sigma(\chi_{c2})&=&\sigma^{(0)}(\chi_{c2})\bigg\{1+\frac{\alpha_s}{\pi}(-9.0312)+\frac{\alpha_s^2}{\pi^2}\bigg[\frac{1}{4}\beta_0\ln\frac{\mu_R^2}{4m^2}(-9.0312)
+2\gamma_{\chi_{c2}}\ln\frac{\mu_\Lambda}{m}\nn\\
&&+\bigg(2.205168(2){ n_H}+4.1844189(5){ n_L}+3.456(3){\rm lbl}-60.504(2)\bigg)
\bigg]\bigg\}
\end{eqnarray}
\end{subequations}
for $m=1.4$ GeV, and
\begin{subequations}\label{cs-total-mc-1.68}
\begin{eqnarray}
\sigma(\chi_{c0})&=&\sigma^{(0)}(\chi_{c0})\bigg\{1+\frac{\alpha_s}{\pi}(2.7728)+\frac{\alpha_s^2}{\pi^2}\bigg[\frac{1}{4}\beta_0
\ln\frac{\mu_R^2}{4m^2}(2.7728)
+2\gamma_{\chi_{c0}}\ln\frac{\mu_\Lambda}{m}\nn\\
&&+\bigg(0.931349(3) n_H -1.423602(1) n_L+2.961(7){\rm lbl}-8.077(2)\bigg)
\bigg]\bigg\}, \\
\sigma(\chi_{c1})&=&\sigma^{(0)}(\chi_{c1})\bigg\{1+\frac{\alpha_s}{\pi}(-3.6598)+\frac{\alpha_s^2}{\pi^2}\bigg[\frac{1}{4}\beta_0
\ln\frac{\mu_R^2}{4m^2}(-3.6598)
+2\gamma_{\chi_{c1}}\ln\frac{\mu_\Lambda}{m}\nn\\
&&+\bigg(0.384512(1) n_H-0.2633821(5) n_L-2.8413(6){\rm lbl}-21.294(1)\bigg)
\bigg]\bigg\},\\
\sigma(\chi_{c2})&=&\sigma^{(0)}(\chi_{c2})\bigg\{1+\frac{\alpha_s}{\pi}(-8.92115)+\frac{\alpha_s^2}{\pi^2}\bigg[\frac{1}{4}\beta_0
\ln\frac{\mu_R^2}{4m^2}(-8.92115)
+2\gamma_{\chi_{c2}}\ln\frac{\mu_\Lambda}{m}\nn\\
&&+\bigg(2.375799(2) n_H+4.6143829(4) n_L+2.4154(5){\rm lbl}-68.447(1)\bigg)
\bigg]\bigg\}
\end{eqnarray}
\end{subequations}
for $m=1.68$ GeV.

In Eqs.~(\ref{cs-total-mc-1.4}) and (\ref{cs-total-mc-1.68}), $\beta_0=(11/3) C_A-(4/3)T_F n_f$ is the one-loop coefficient of the QCD $\beta$ function, with $T_F=\tfrac{1}{2}$ and $n_f$ signifying the number of active quark flavors. In this work, we take $n_f=n_L+n_H$, where $n_L=3$
labels the number of light quark flavors and $n_H=1$ indicates the number of heavy quark flavors.
In addition, the symbol `lbl' labels the contributions from the Feynman diagrams with the ``light-by-light'' topology,
which are illustrated by the last diagram in Fig.~\ref{fig-feynman-diagram}.
Note that the occurrence of $\gamma_{\chi_{cJ}}\ln \mu_\Lambda^2$ terms is reminiscent of the remnant of the uncancelled single IR pole,
while the occurrence of the $\beta_0 \ln \mu_R^2$ simply reflects the renormalization group invariance.

\section{Phenomenology\label{sec-phen}}

In this section, we apply the formulas obtained in section~\ref{sec-calculation} to make a concrete
phenomenological analysis.
Relativistic corrections for charmonium are expected to bear a magnitude of 30\%.
Although the relativistic corrections for $e^+e^-\to \chi_{cJ}+\gamma$ have recently been thoroughly investigated
in Ref.~\cite{Brambilla:2017kgw}, the corresponding uncertainties appear to be substantial,
because the $\mathcal{O}(v^2)$ NRQCD matrix elements are poorly constrained.
For simplicity we choose to neglect this sort of contribution in our phenomenological analysis.

First we need to fix the various input parameters.
We take the running QED coupling constant evaluate at the $B$ factory energy scale, $\alpha(\sqrt{s})=1/130.9$.
The QCD running coupling constant is evaluated to two-loop accuracy with the aid of the package
{\tt RunDec}~\cite{Chetyrkin:2000yt}. The NRQCD LDME for $\chi_{cJ}$ is
approximated by the first derivative of the Schro\"{o}dinger radial wave function at origin through
\begin{eqnarray}\label{eq-ldme-wave-funciton}
\langle \mathcal{O}(^3P_J) \rangle\approx\frac{3N_c}{2\pi}|R^\prime_{1P}(0)|^2.
\end{eqnarray}
The $1P$ radial wave function at origin for $\chi_c$ varies with different quark potential models. For instance,
$|R^\prime_{1P}(0)|^2=0.075\;{\rm GeV}^5$ in Buchm\"uller-Tye (BT) potential model, and $|R^\prime_{1P}(0)|^2=0.1296\; {\rm GeV}^5$ in Cornell potential
model~\cite{Eichten:1995ch,Eichten:2019hbb}.
Substituting these values into (\ref{eq-ldme-wave-funciton}), one immediately obtains the corresponding NRQCD LDME
$\langle \mathcal{O}(^3P_J)\rangle=0.107 \; {\rm GeV}^5$ from BT potential model, and $\langle \mathcal{O}(^3P_J)\rangle=0.186 \; {\rm GeV}^5$ from
Cornell model~\footnote{In Ref.~\cite{Chung:2008km},
the LDME $\langle \mathcal{O}(^3P_J)\rangle$ is fitted via equating the NRQCD factorization
predictions accurate through ${\cal O}(\alpha_s)$ with the measured values for $\chi_{c0,2}\to 2\gamma$ compiled by the particle data group~\cite{Tanabashi:2018oca}. The LDME is determined to be $\langle \mathcal{O}(^3P_J)\rangle=0.060^{+0.043}_{-0.029} \; {\rm GeV}^5$~\cite{Chung:2008km}, which seems to be considerably
smaller than the values given by potential models. Nevertheless, the $\mathcal{O}(\alpha_s^2)$ corrections $\chi_{c0,2}\to \gamma\gamma$
turns out to be substantial~\cite{Sang:2015uxg}.
For consistency, we will not use the fitted value of LDME in \cite{Chung:2008km} in current work.}.

By setting the renormalization scale $\mu_R=\sqrt{s}/2$ and the NRQCD factorization scale $\mu_\Lambda=m$,
we can express the cross sections of $e^+e^-\to \chi_{cJ}+\gamma$ as the following power series:
\begin{subequations}\label{corrections-1-mc-1.4}
\begin{eqnarray}
\sigma(\chi_{c0}+\gamma)&=&\sigma^{(0)}(\chi_{c0}+\gamma)\; \left[1+0.62\alpha_s-0.28\alpha_s^2+\mathcal{O}(\alpha_s^3)\right],\\
\sigma(\chi_{c1}+\gamma)&=&\sigma^{(0)}(\chi_{c1}+\gamma)\; \left[1-1.01\alpha_s-3.21\alpha_s^2+\mathcal{O}(\alpha_s^3)\right],\\
\sigma(\chi_{c2}+\gamma)&=&\sigma^{(0)}(\chi_{c2}+\gamma)\; \left[1-2.87\alpha_s-6.71\alpha_s^2+\mathcal{O}(\alpha_s^3)\right]
\end{eqnarray}
\end{subequations}
for $m=1.40$ GeV, and
\begin{subequations}\label{corrections-1-mc-1.68}
\begin{eqnarray}
\sigma(\chi_{c0}+\gamma)&=&\sigma^{(0)}(\chi_{c0}+\gamma)\;  \left[1+0.88\alpha_s-0.33\alpha_s^2+\mathcal{O}(\alpha_s^3)\right],\\
\sigma(\chi_{c1}+\gamma)&=&\sigma^{(0)}(\chi_{c1}+\gamma)\; \left[1-1.16\alpha_s-3.19\alpha_s^2+\mathcal{O}(\alpha_s^3)\right],\\
\sigma(\chi_{c2}+\gamma)&=&\sigma^{(0)}(\chi_{c2}+\gamma)\; \left[1-2.84\alpha_s-6.76\alpha_s^2+\mathcal{O}(\alpha_s^3)\right]
\end{eqnarray}
\end{subequations}
for $m=1.68$ GeV.
From Eqs.~(\ref{corrections-1-mc-1.4}) and (\ref{corrections-1-mc-1.68}), we clearly observe that the $\mathcal{O}(\alpha_s^2)$ corrections
are less important than the $\mathcal{O}(\alpha_s)$ corrections. Therefore, to some extent,
the perturbative expansion in $\alpha_s$ exhibits a decent convergence behavior, especially for the $\chi_{c0}+\gamma$ channel,
in which the NNLO corrections only plays a minor role.

Assuming the LDME $\langle \mathcal{O}(^3P_J)\rangle=0.107 \; {\rm GeV}^5$ as given by the BT potential model,
and adopting two benchmark values of charm quark mass,
in Table~\ref{table-numerical-prediction-BT} we tabulate the NRQCD predictions to production rates for $e^+e^-\to \chi_{cJ}+\gamma$
at various level of perturbative accuracy.
The uncertainties affiliated with the cross sections are estimated by varying $\mu_R$ from $2m$ to $\sqrt{s}$,
with the central values evaluated at $\mu_R=\sqrt{s}/2$.
We notice the cross sections with $m=1.68$ GeV are considerably smaller than those with $m=1.4$ GeV, which can be attributed to the
factor $1/m^3$ that arise in the NRQCD SDCs for $\sigma(e^+e^-\to \chi_{cJ}+\gamma)$, as is evident in \eqref{sdcs-0}.
Therefore we expect that the uncertainty due to charm quark pole mass is considerably greater
than that from varying the renormalization as well as the NRQCD factorization scales.
For more discussion about the heavy quark pole mass, we refer the interested readers to Refs.~\cite{Marquard:2015qpa,Kataev:2015gvt,Ayala:2019hkn,Mateu:2017hlz}.

It is interesting to note that, the perturbative corrections to $\sigma(e^+e^-\to \chi_{c2}+\gamma)$, {\it i.e.},
both the NLO and NNLO corrections, are sizable and negative.
Incorporating the NNLO corrections reduces the LO prediction by almost one order of magnitude.
In contrast, both the $\mathcal{O}(\alpha_s)$ and $\mathcal{O}(\alpha_s^2)$ corrections are moderate for $\chi_{c1}+\gamma$ production, and
have minor effect for the $\chi_{c0}+\gamma$. In addition, from Table~\ref{table-numerical-prediction-BT} one may also observe that
the production rates for $\chi_{c0,1}+\gamma$ are insensitive to the factorization scale $\mu_\Lambda$.

\begin{table}
\caption{NRQCD predictions to $\sigma(\chi_{cJ}+\gamma)$ at various levels of accuracy in $\alpha_s$ at $B$ factory.
The LDME $\langle \mathcal{O}(^3P_J)\rangle=0.107 \; {\rm GeV}^5$ is taken from BT potential model.
The errors are estimated by sliding the renormalization scale $\mu_R$ from $2m$ to $\sqrt{s}$.}
\begin{center}
\label{table-numerical-prediction-BT}
\begin{tabular}{|c|c|c|c|c|c|}
\hline
\multicolumn{5}{|l|}{$m=1.40$ GeV} \\
\hline
 & \multicolumn{2}{|c|}{} &\multicolumn{1}{|c|}{$\mu_\Lambda=1$ GeV} &  \multicolumn{1}{|c|}{$\mu_\Lambda=m$ }\\
\cline{2-5}
 \diagbox{$\chi_{cJ}$}{$\sigma$ (fb)}{Order}  & LO & NLO  & NNLO   & NNLO  \\
\hline
$\chi_{c0}+\gamma$ & $2.52$ & $2.83^{+0.06}_{-0.04}$ & $2.96^{+0.05}_{-0.04}$  & $2.82^{+0.01}_{-0.03}$ \\
\hline
$\chi_{c1}+\gamma$ & $25.96$ & $20.72^{+0.75}_{-1.05}$ & $17.91^{+0.89}_{-1.21}$  & $16.83^{+1.20}_{-1.79}$\\
\hline
$\chi_{c2}+\gamma$ & $10.02$ & $4.24^{+0.83}_{-1.16}$ & $1.34^{+0.92}_{-1.23}$  & $1.03^{+1.01}_{-1.40}$ \\
\hline
\multicolumn{5}{|l|}{$m=1.68$ GeV} \\
\hline
$\chi_{c0}+\gamma$ & $1.18$ & $1.39^{+0.03}_{-0.03}$ & $1.48^{+0.03}_{-0.03}$  & $1.38^{+0.01}_{-0.01}$ \\
\hline
$\chi_{c1}+\gamma$ & $15.98$ & $12.25^{+0.54}_{-0.50}$ & $10.87^{+0.45}_{-0.37}$  & $9.84^{+0.75}_{-0.72}$\\
\hline
$\chi_{c2}+\gamma$ & $6.60$ & $2.84^{+0.54}_{-0.50}$ & $1.03^{+0.58}_{-0.52}$  & $0.71^{+0.67}_{-0.63}$\\
\hline
\end{tabular}
\end{center}
\end{table}

Very recently, {\tt Belle} experiment measured $\sigma(e^+e^-\to \chi_{c1}+\gamma)= 17.3^{+4.2}_{-3.9}({\rm stat.})\pm 1.7 ({\rm syst.})\;{\rm fb}$, yet failed to observe $\chi_{c0,2}+\gamma$~\cite{Jia:2018xsy} events.
It is remarkable that our prediction to $\chi_{c1}+\gamma$ production rate with $m=1.4$ GeV is compatible with {\tt Belle} measurement!
From Table~\ref{table-numerical-prediction-BT}, we also notice that the production rates for $\chi_{c0,2}+\gamma$ are
roughly one order-of-magnitude smaller than that for $\chi_{c1}+\gamma$,
which probably explains why $\chi_{c0,2}+\gamma$ have remained undiscovered by {\tt Belle} experiment.

Were the value of LDME chosen from Cornell model instead of BT model, our predicted cross sections would be enhanced roughly by
a factor of 1.7. The NRQCD predictions to the cross sections with $m=1.4$ GeV generally overshoot
the upper bound of {\tt Belle} measurement. Nevertheless, it is interesting to note that the
NNLO prediction to $\sigma(\chi_{c1}+\gamma)$ with $m=1.68$ GeV is about $17.01^{+1.30}_{-1.25}$ fb,
in perfect agreement with the {\tt Belle} measurement within error.

\begin{figure}[htbp]
 	\centering
 \includegraphics[width=0.45\textwidth]{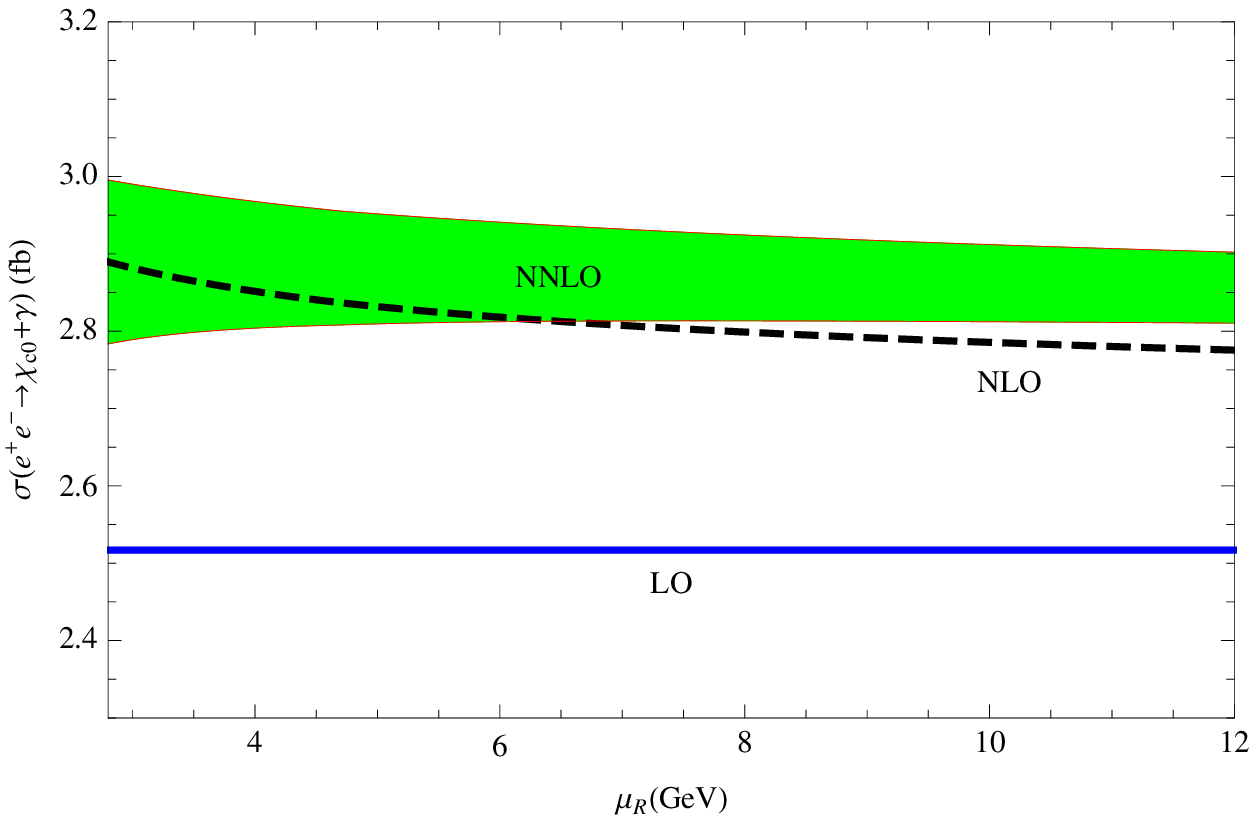}
    \includegraphics[width=0.45\textwidth]{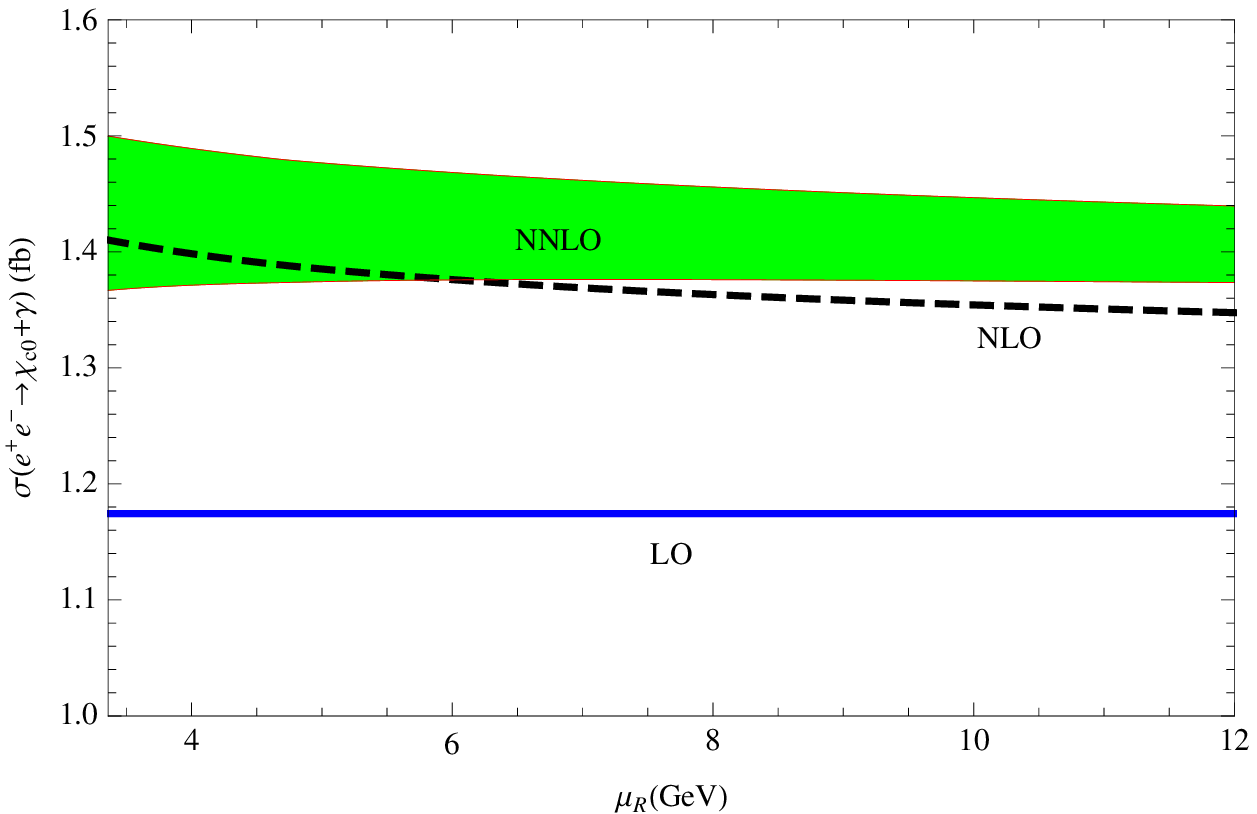}
 \includegraphics[width=0.45\textwidth]{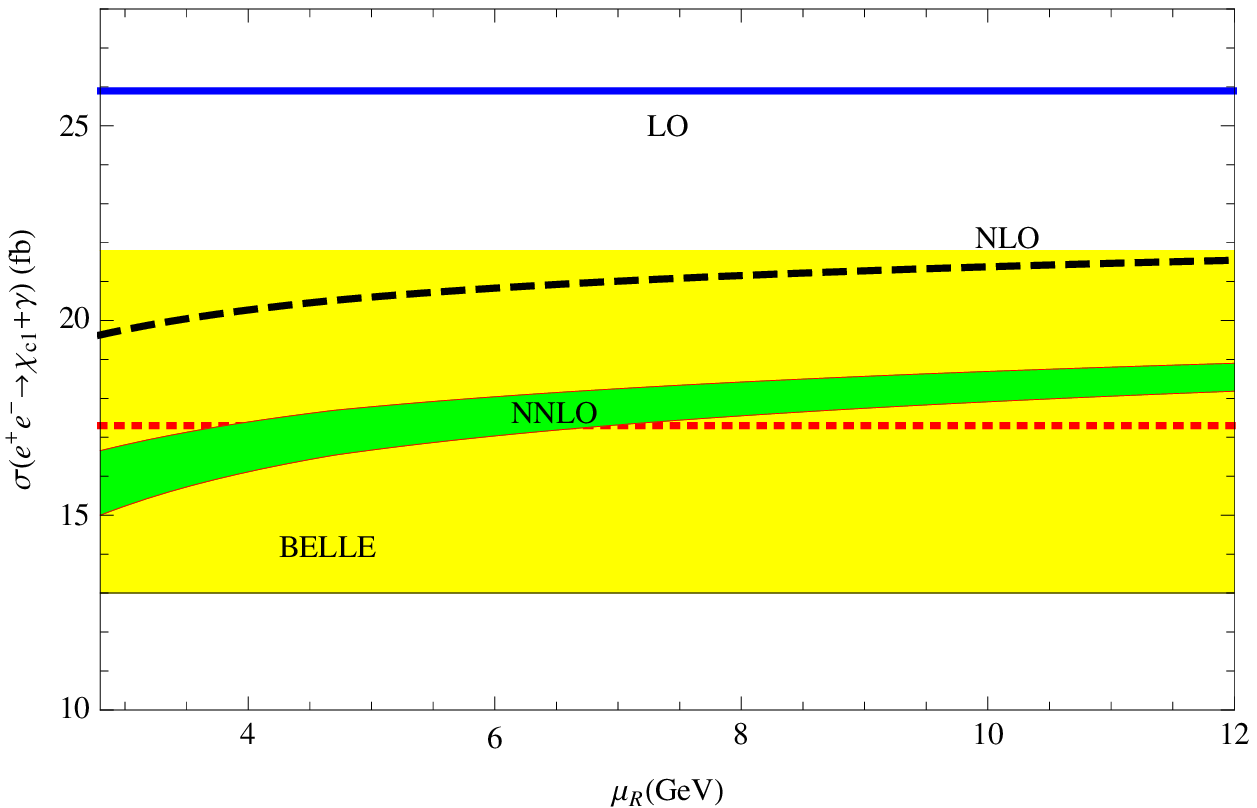}
  \includegraphics[width=0.45\textwidth]{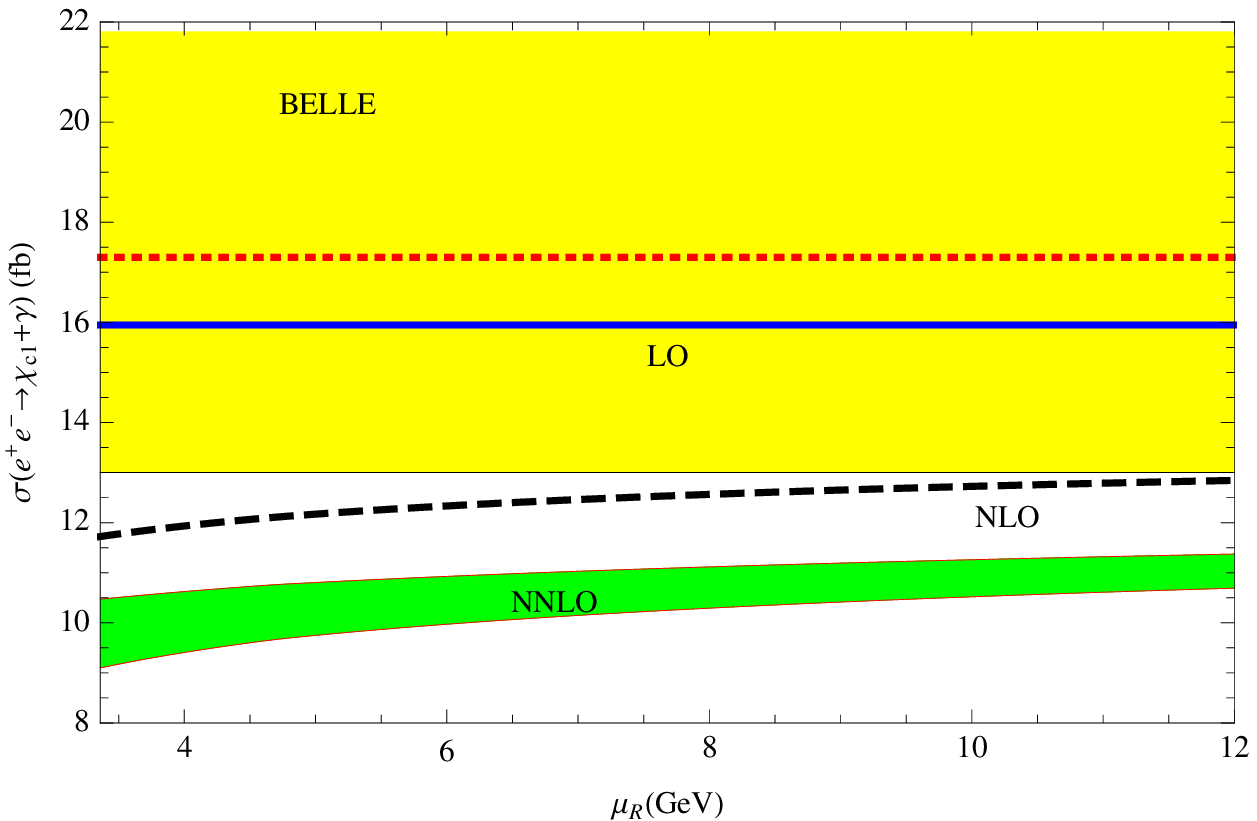}
  \includegraphics[width=0.45\textwidth]{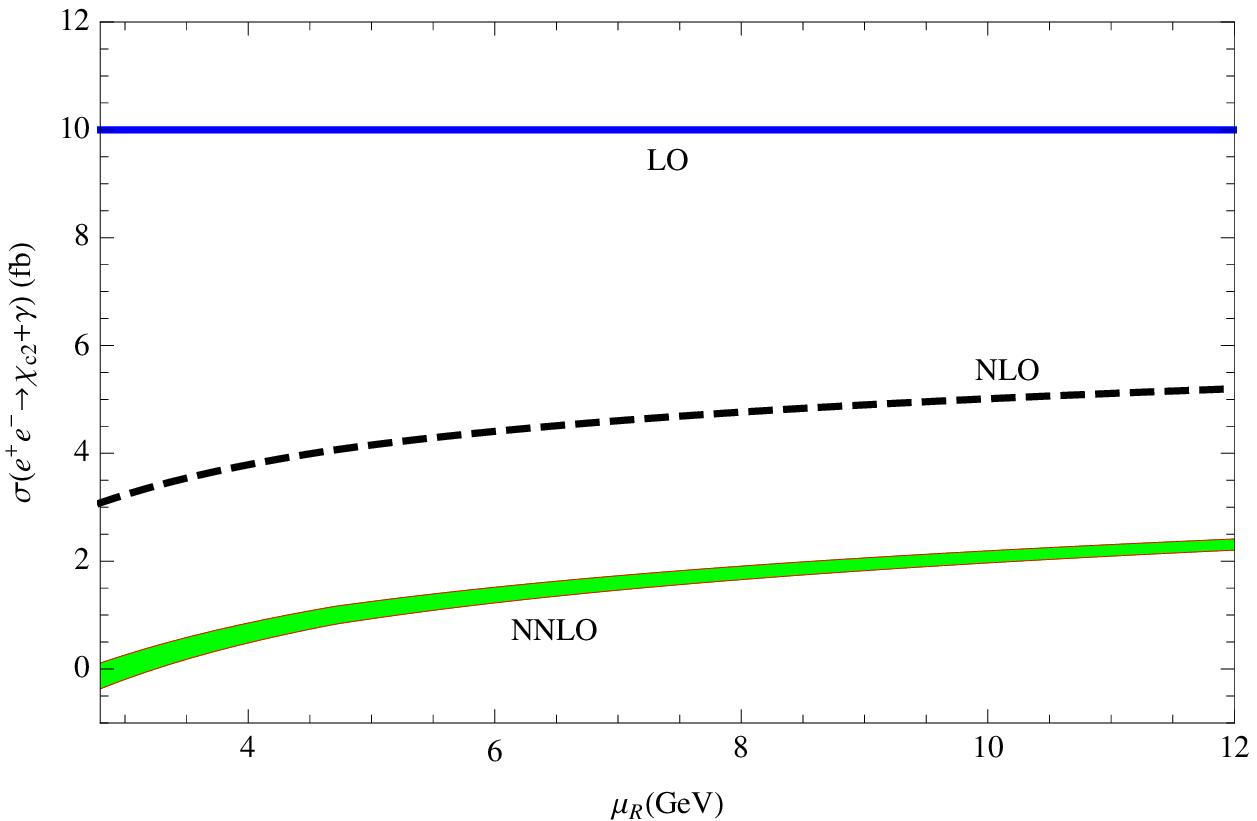}
  \includegraphics[width=0.45\textwidth]{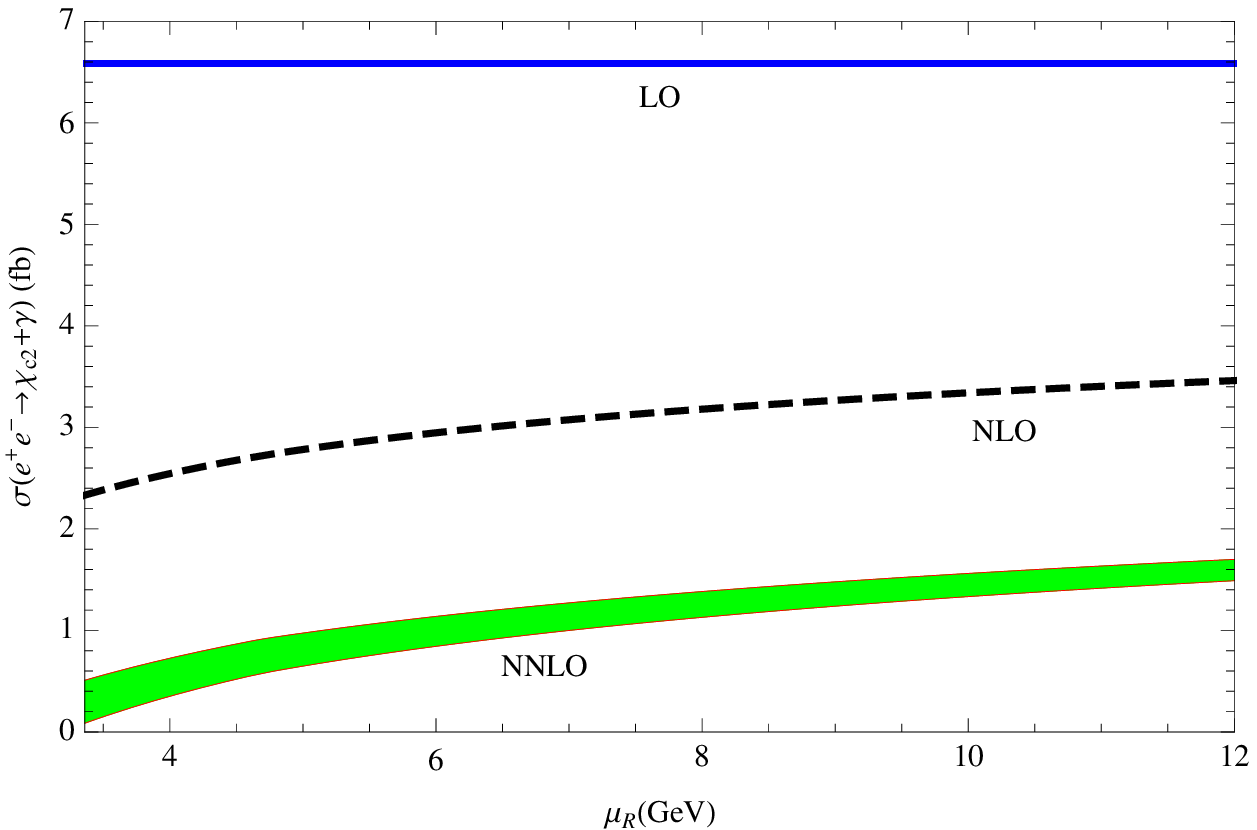}
 \caption{NRQCD predictions for $\sigma(e^+e^-\to \chi_{cJ}+\gamma)$ as a function of $\mu_R$ at various levels of accuracy in $\alpha_s$,
 yet at the lowest order in $v$. The labels ``LO", ``NLO" and ``NNLO" represent the contributions from $\mathcal{O}(\alpha_s^0)$, $\mathcal{O}(\alpha_s^0 )$+$\mathcal{O}(\alpha_s^1)$, and $\mathcal{O}(\alpha_s^0)$+$\mathcal{O}(\alpha_s^1)$+$\mathcal{O}(\alpha_s^2)$, respectively.
 The value of the LDME is taken from the BT potential model.
 We take $m=1.4$ GeV for the figures in the left panel,
 while $m=1.68$ GeV in the right panel. The green band represents the uncertainty band by varying $\mu_\Lambda$ from 1 GeV to $m$.
 In addition, we also display the {\tt Belle} measurement by the yellow band for the $e^+e^-\to \chi_{c1}+\gamma$ channel.
 \label{fig-mu-dependence}}
 \end{figure}

In Figure~\ref{fig-mu-dependence}, we plot the cross sections as a function of the renormalization scale $\mu_R$ at various level of
perturbative accuracy, with the value of LDME taken from the BT potential model. We take $m=1.4$ GeV on the left panel, and
$m=1.68$ GeV on the right panel. The green band labeled with ``NNLO" is obtained by varying the factorization scale from 1 GeV to $m$.
In order to facilitate comparison, we also demonstrate the {\tt Belle} data by the yellow band in the plot of $\chi_{c1}+\gamma$,
where the red-dotted curve corresponds to the central value of the experimental measurement.
We observe that the NNLO prediction seems to have a slightly reduced $\mu_R$-dependence relative to the NLO prediction.
More importantly, our NRQCD prediction to $\sigma(\chi_{c1}+\gamma)$ with $m=1.4$ GeV agrees well with the {\tt Belle} measurement
within the reasonable range of $\mu_R$.

\section{Summary}
\label{summary}

In summary, in this work we have computed the $\mathcal{O}(\alpha_s^2)$ corrections to $\sigma(e^+e^-\to \chi_{cJ}+\gamma)$ at $B$ factory.
We choose two benchmark values for charm quark mass $m=1.4$ GeV and $m=1.68$ GeV,
which correspond to the one-loop and two-loop charm quark pole mass, respectively.
For the first time, we have verified that NRQCD factorization remains valid for exclusive $P$-wave quarkonium production to two-loop
order.
The impact of NNLO perturbative corrections is found to be significant for $\chi_{c2}+\gamma$, moderate for $\chi_{c1}+\gamma$,
and rather marginal for $\chi_{c0}+\gamma$.
Unlike the electromagnetic decays $\chi_{c0,2}\to\gamma\gamma$, the NNLO perturbative corrections to $\sigma(e^+e^-\to \chi_{cJ}+\gamma)$
are found to be smaller than the NLO pertubative corrections for all $\chi_{cJ}+\gamma$ channels, which may indicate a satisfactory
convergence in perturbative expansion.
The NRQCD LDMEs are approximated by the first derivative of the $\chi_{cJ}$
wave function at the origin deduced from nonrelativistic potential models.
When taking  the BT potential model as input, we find the NNLO predictions to the cross section of $\chi_{c1}+\gamma$ with $m=1.4$ GeV is consistent with the {\tt Belle} measurement within errors. On the other hand, when taking the LDME from Cornell model,
the NRQCD prediction with $m=1.68$ GeV turns out to be also consistent with the experimental measurement.
After including the NNLO perturbative corrections, the production rates for $\chi_{c0,2}+\gamma$ are still found to be
much smaller than that of $\chi_{c1}+\gamma$,
which may naturally explain why the $e^+e^-\to \chi_{c0,2}+\gamma$ channels have escaped the experimental detection until today.

\section*{Acknowledgement}
The work of W.-L. S. is supported by the National Natural Science Foundation
of China under Grants No. 11975187 and the Natural Science Foundation of ChongQing under Grant No. cstc2019jcyj-msxm2667.
The work of F.~F. is supported by the National Natural
Science Foundation of China under Grant No. 11875318,
No. 11505285, and by the Yue Qi Young Scholar Project
in CUMTB.
The work of Y.~J. is supported in part by the National Natural Science Foundation of China
under Grants No.~11925506, 11875263,
No.~11621131001 (CRC110 by DFG and NSFC).



\end{document}